\pdfoutput=1
\documentclass[10pt,nocopyrightspace,preprint]{sigplanconf}

\newfont{\ttlfnt}{phvb8t at 18pt}
\newfont{\aufnt}{phvr8t at 12pt}
\newfont{\affaddr}{phvr8t at 9pt}
\let\sigplantitle\title
\renewcommand{\title}[1]{\sigplantitle{\setlength{\baselineskip}{22pt}\ttlfnt #1}}
\let\sigplanauthor\authorinfo
\renewcommand{\authorinfo}[3]{\sigplanauthor{\aufnt
    #1}{\setlength{\baselineskip}{12pt}\affaddr #2}{#3}}

\makeatletter
\def \@maketitle {%
  \begin{center}
  \@settitlebanner
  \let \thanks = \titlenote
  \vskip -5pt
  {\leftskip = 0pt plus 0.25\linewidth
   \rightskip = 0pt plus 0.25 \linewidth
   \parfillskip = 10pt
   \spaceskip = .7em
   \noindent \bfseries \@titletext \par}
  \vskip 6pt
  \noindent \Large \@subtitletext \par
  \vskip 12pt
  \ifcase \@authorcount
    \@latex@error{No authors were specified for this paper}{}\or
    \@titleauthors{i}{}{}\or
    \@titleauthors{i}{ii}{}\or
 \fi
  \vspace{-1.5pc}
  \end{center}}
\makeatother

\newif\ifnmd
\InputIfFileExists{user-is-nathan-dunfield}{\nmdtrue}{\nmdfalse}

\usepackage{amsmath, amsthm, amssymb}
\usepackage{fourier} 

\ifnmd   
\fi

\usepackage{graphicx}
\usepackage{paralist}
\usepackage{color}
\usepackage[colorlinks,linkcolor=black,bookmarksopen, bookmarksnumbered,citecolor=black,urlcolor=black]{hyperref}

\def\RCS$#1: #2 ${\expandafter\def\csname RCS#1\endcsname{#2}}
\RCS$Revision: 11167 $
\RCS$Date: 2011-03-23 12:42:04 -0500 (Wed, 23 Mar 2011) $

\graphicspath{{figs/}}

\swapnumbers

\def\R{\mathbb{R}}

\def\Z{\mathbb{Z}}
\def\F{\mathbb{F}}

\DeclareMathOperator{\boundary}{\partial}

\def\P{\mathcal{P}}

\newcommand{\abs}[1]{\lvert#1\rvert}
\newcommand{\norm}[1]{\lVert#1\rVert}

\newcommand{\hyp}{\nobreakdash-\hspace{0pt}}
\newcommand{\3}[1]{3\hyp}
\newcommand{\one}[1]{1\hyp}
\newcommand{\two}[1]{2\hyp}

\newcommand{\Area}{\mathrm{Area}}
\newcommand{\NP}{\mathbf{NP}}

\newcommand{\coNP}{\mbox{\bf co-NP}}
\newcommand{\maps}{\colon\thinspace}
\newcommand{\cF}{\mathcal F}
\newcommand{\N}{\mathbb N}
\newcommand{\pair}[1]{\left\langle #1 \right\rangle}

\newcommand{\setdefm}[3]{{  #1\{  {#2}  \ #1 | \   {#3}  #1\} }}

\newcommand{\kernoverline}[3]{{\mkern #1mu\overline{\mkern-#1mu #2\mkern-#3mu}\mkern#3mu}}

\newcommand{\ubar}{{\kernoverline{3.5}{u}{1.5}}}

\usepackage[all,import,rotate]{xy}
\newenvironment{xyoverpic}[3]
{%
\begin{xy}
\xyimport#1{\includegraphics[#2]{#3}}
}{\end{xy}}

\newenvironment{cxyoverpic}[3]
{%
\begin{center}
\centering\leavevmode\small
\begin{xyoverpic}{#1}{#2}{#3}
}{\end{xyoverpic}
\end{center}}

\newtheorem{theorem}{Theorem}[section]
\newtheorem*{theorem*}{Theorem}

\newtheorem*{proposition*}{Proposition}
\newtheorem{lemma}[theorem]{Lemma}
\newtheorem*{lemma*}{Lemma}

\newtheorem*{claim*}{Claim}

\newtheorem*{axiom*}{Axiom}
\newtheorem{conjecture}[theorem]{Conjecture}
\newtheorem*{conjecture*}{Conjecture}
\newtheorem{corollary}[theorem]{Corollary}
\newtheorem*{corollary*}{Corollary}

\theoremstyle{definition}

\newtheorem*{definition*}{Definition}

\newtheorem*{example*}{Example}

\newtheorem*{exercise*}{Exercise}
\newtheorem*{recall*}{Recall}

\theoremstyle{remark}

\newtheorem*{note*}{Note}
\newtheorem{remark}[theorem]{Remark}
\newtheorem*{remark*}{Remark}

\newtheorem*{notation*}{Notation}
\newtheorem*{question*}{Question}

\newtheorem*{fact*}{Fact}

\theoremstyle{definition}

\theoremstyle{remark}

\theoremstyle{plain}
\newtheorem{knotgenus}[theorem]{Knot Genus}
\newtheorem{minarea}[theorem]{Least Spanning Area}
\newtheorem{ROBCP}[theorem]{Relative OBCP}
\newtheorem{goal}[theorem]{Problem}
\newtheorem{obcpd}[theorem]{OBCP-D}

\makeatletter
  \let\c@theorem=\c@subsection
  \let\c@figure=\c@subsection
  \let\p@figure=\p@subsection
 
 \let\cl@equation=\cl@subsection
  \let\c@table=\c@subsection
  \let\c@equation=\c@subsection
  \let\p@equation=\p@subsection
 
 \let\cl@equation=\cl@subsection
\makeatother

\definecolor{darkgrn}{rgb}{0, 0.75, 0}

\begin{document}
\conferenceinfo{SCG'11,} {June 13--15, 2011, Paris, France.} 
\CopyrightYear{2011} 
\crdata{978-1-4503-0682-9/11/06} 

\ifnmd
\fi

\title{The Least Spanning Area of a Knot \\ and 
the Optimal Bounding Chain Problem}

%

\authorinfo{Nathan M. Dunfield}
{University of Illinois, Mathematics \\ 
1409 W. Green St. \\ 
Urbana IL, 61801, USA} 
{}
\authorinfo{Anil N. Hirani}
{University of Illinois, Computer Science \\ 
201 N. Goodwin Ave. \\ 
Urbana IL, 61801, USA}
{}

\maketitle

\begin{abstract}
  Two fundamental objects in knot theory are the minimal genus surface
  and the least area surface bounded by a knot in a \3-dimensional
  manifold.  When the knot is embedded in a general 3-manifold, the
  problems of finding these surfaces were shown to be $\NP$-complete
  and $\NP$-hard respectively. However, there is evidence that the
  special case when the ambient manifold is $\mathbb{R}^3$, or more
  generally when the second homology is trivial, should be
  considerably more tractable.  Indeed, we show here that a natural
  discrete version of the least area surface can be found in
  polynomial time.

  The precise setting is that the knot is a 1-dimensional subcomplex
  of a triangulation of the ambient 3-manifold.  The main tool we use
  is a linear programming formulation of the Optimal Bounding Chain
  Problem (OBCP), where one is required to find the smallest norm
  chain with a given boundary. While the decision variant of OBCP is
  $\NP$-complete in general, we give conditions under which it can be
  solved in polynomial time.  We then show that the least area surface
  can be constructed from the optimal bounding chain using a standard
  desingularization argument from 3-dimensional topology.

  We also prove that the related Optimal Homologous Chain Problem is
  $\NP$-complete for homology with integer coefficients, complementing
  the corresponding result of Chen and Freedman for mod 2 homology.
\end{abstract}

\section{Introduction} 
\label{sec-intro}

A knot $K$ is a simple closed loop in an ambient \3-dimensional
manifold $Y$.  Provided $K$ is null-homologous, which is always the
case if $Y = \R^3$, there is an embedded orientable surface $S$ in $Y$
whose boundary is $K$ (equivalently $S$ is a compact smooth orientable
surface in $Y$ without self-intersections and with boundary $K$). A
fundamental property of $K$ is the minimal genus of such an $S$, which
is denoted $g(K)$ (we take $g(K) = \infty$ if there are no such
surfaces).  In the 1960s, Haken used normal surface theory to give an
algorithm for computing $g(K)$, opening the door to a whole subfield
of low-dimensional topology and leading to the discovery of algorithms
for determining a wide range of topological properties of \3-manifolds
\cite{MatveevAlgorithmsBook}.  However, algorithms based on normal
surface theory are quite slow in practice \cite{Burton2010b,
  Burton2010a, Burton2010c}, and there are very few results that have
been verified via such normal surface computations
\cite{DunfieldGaroufalidis2010, BurtonRubinsteinTillmann2010}.
Moreover, in some cases the underlying problems have been shown to be
fundamentally difficult.  In their foundational work, Agol, Hass, and
Thurston showed that the following decision problem is $\NP$--complete
\cite{AgolHassThurston}:
\begin{knotgenus}
  Given an integer $g_0$ and a knot $K$ embedded in the 1-skeleton of
  a triangulation of a closed \3-manifold $Y$, is $g(K) \leq g_0$?
\end{knotgenus}

\noindent
While Knot Genus is $\NP$--complete, when $Y$ is orientable and the
second Betti number $b_2(Y) = \mathrm{rank} \,\big( H_2(Y; \Z) \big)$
is $0$, for instance $Y = S^3$, then this problem likely simplifies.
While their project is not yet complete, Agol, Hass, and Thurston have
developed a very promising approach to showing that when $b_2(Y) = 0$
there is a certificate for the complementary problem $g(K)
\geq g_0$ which can be verified in polynomial time.  This would mean
that this special case of Knot Genus is also in $\coNP$, raising the
possibility of a polynomial-time algorithm when $b_2(Y) = 0$.
However, currently there are no known algorithms which exploit the
fact that $b_2(Y) = 0$.  Despite this, our long-term goal is
\begin{conjecture}\label{conj-knot-P}
  For orientable $Y$ with $b_2(Y) = 0$ the Knot Genus problem is in
  \textbf{P}.
\end{conjecture}

Here, we study the related problem of finding the least area surface
bounded by a knot.  This problem has its origin in classical
differential geometry, as we now sketch starting with the case where
the ambient manifold $Y$ is $\R^3$.  For a smooth knot $K$ in $\R^3$
there is always a smooth embedded orientable surface $S \subset \R^3$
with $\partial S = K$.  By deep theorems in Geometric Measure Theory,
there always exists such a surface $S_0$ of least area
\cite{MorganGMTBook}.  A least area surface $S_0$ is necessarily
\emph{minimal} in that it has mean curvature 0 everywhere, like the
surface of a soap-bubble.  It is typically impossible to find the
least area surface analytically, and the first paper on numerical
methods for approximating $S_0$ appeared in 1927 \cite{Douglas1927}.
An algorithm to deal with arbitrary $K$ was first given by Sullivan
\cite{SullivanThesis} in 1990; see also \cite{Parks1986, Parks1992,
  ParksPitts1997, PinkallPolthier1993, Polthier2005} for alternate
approaches and numerical experiments.

Of course, one can consider this question for null-homologous knots in
an arbitrary Riemannian \3-manifold $Y$, and one has the same
existence theorems for least area surfaces when $Y$ is closed.  Agol,
Hass, and Thurston considered a certain discrete version of this
problem, and showed that the question of whether $K$ bounds a surface
of area $\leq A_0$ is $\NP$-hard \cite{AgolHassThurston}.  Because
they put no restriction on the surface involved, it is not clear if
their question is in $\NP$.  Here, we consider another discretization,
the Least Spanning Area Problem of Section~\ref{sec-least-area}, which
is a little more combinatorial and will thus turn out to be
$\NP$-complete (Theorem~\ref{thm-LSA-hard}).  For this problem, we
show
\begin{theorem}\label{thm-min-area}
  For orientable $Y$ with $b_2(Y) = 0$ the Least Spanning Area Problem
  is in \textbf{P}.
\end{theorem}
\noindent
In Section~\ref{sec-future}, we discuss how the ideas behind
Theorem~\ref{thm-min-area} might be used to attack
Conjecture~\ref{conj-knot-P}, as these two questions have a very
similar flavor.

One of two key tools behind Theorem~\ref{thm-min-area} is the
following type of combinatorial optimization problem.  For a finite
simplicial complex $X$, fix an $\ell^1$-norm on the simplicial chains
$C_*(X ; \Z)$ by assigning each simplex an arbitrary nonnegative
weight. (For example, every simplex could have weight 1, or if $X$ is
a geometric mesh in $\R^3$ one could take each weight to be the
length/area/volume of the simplex itself.)  For a chain $a \in C_n(X ;
\Z)$ the Optimal Homologous Chain Problem (OHCP) is to find a chain $c
\in C_n(X ; \Z)$ homologous to $a$ with $\norm{c}_1$ minimal.  As
there are many choices for $c$, this might seem like a hard
problem. Indeed, consider the decision problem variant, OHCP-D: given $X$, $a$, and
$L \in \N$ is there a $c$ homologous to $a$ with $\norm{c}_1 \le L$? We show:
\begin{theorem}\label{thm-OHCP-NP}
  The OHCP-D with integer coefficients is \textbf{NP}-complete.  
\end{theorem}
\noindent
Chen and Freedman established the same result when one uses homology
with $\F_2$-coefficients \cite{ChFr2010aalt}.  However, with the
addition of a simple condition on $X$ (which often holds in geometric
applications), Dey, Hirani, and Krishnamoorthy \cite{DeHiKr2010} have
used linear programming to solve the OHCP for $\Z$ coefficients in
polynomial time, and this will be a key tool here.

We also need to consider the related Optimal Bounding Chain Problem
(OBCP): given $b \in C_{n-1}(X; \Z)$, find the $c \in C_{n}(X; \Z)$
where $\partial_n c = b$ and $\norm{c}_1$ is minimal.  (Of course,
this is only interesting when $[b]$ is $0$ in $H_{n-1}(X; \Z)$ as
otherwise no such $c$ exists.)  If $H_n(X)=0$ then the OBCP is
equivalent to a related instance of the OHCP
(Theorem~\ref{thm-OBCP-OHCP-equiv}).  Thus one can often use the
method of \cite{DeHiKr2010} to solve such problems quickly
(Cor.~\ref{thm-OBCP-P}).  Conversely, we describe how a key
construction in \cite{AgolHassThurston} shows that the decision
problem variant of OBCP is $\NP$-complete in general
(Theorem~\ref{thm-obcp-np}), and we then modify the construction to
prove Theorem~\ref{thm-OHCP-NP}.

To prove Theorem~\ref{thm-min-area}, we reduce the Least Area Surface
problem to the OBCP via a standard desingularization method from
3-dimensional topology that turns an arbitrary 2-cycle into an
embedded surface that is homologous to it.  This surface can be built
constructively, and we outline in Section~\ref{sec-future} how this
gives an approach to Conjecture~\ref{conj-knot-P}.  As a preview to
this material, we give an simple application of the OBCP in
Section~\ref{sec-toy} to the toy problem of finding the shortest path
between opposite sides of a triangulated square.

\section{Optimal chain problems}
\label{sec:OBCP}

\begin{figure}[t]
 \includegraphics[scale = 0.2]{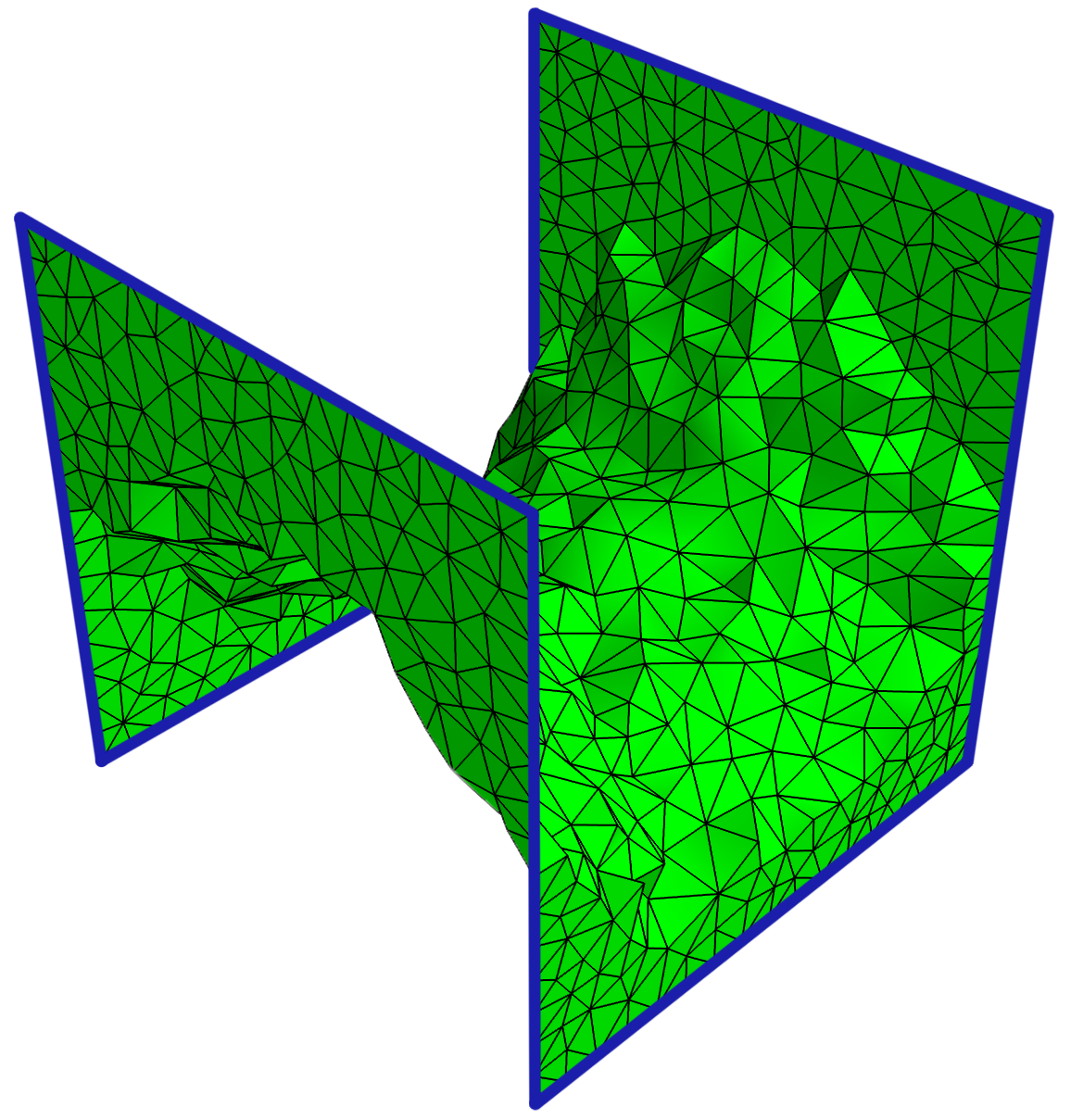}
 \caption{ The surface shown is the
   solution to the OBCP for the 1-cycle $b$ shown in dark blue.  Here
   $X$ is a cube triangulated with 19,201
   tetrahedra, 39,758 triangles, 24,256 edges and 3,700 vertices, and
   the 1-cycle $b$ is a Hamiltonian cycle of the cube corners.  The
   surface approximates Scherk's minimal surface.
 }
  \label{fig-scherk}
\end{figure}

For the rest of this paper, all homology will be over $\Z$, and so we
drop the coefficients from the notation.  As in the introduction, we
consider a finite simplicial complex $X$ with an $\ell^1$-norm on
$C_*(X)$, and recall the Optimal Homologous Chain Problem (OHCP):
given $a \in C_n(X)$, minimize $\norm{c}_1$ over all $c = a
+ \partial_{n+1} x$ with $x \in C_{n + 1}(X)$.  (Here $a$ need not be
a cycle.)  The framework of \cite{DeHiKr2010} is that minimizing
$\norm{c}_1$ can be reformulated as minimizing some linear functional
over the lattice points in a convex region defined by linear
inequalities, i.e.~an integer linear programming problem.  While
integer linear programming is $\NP$-complete, when the matrix for the
boundary map $\partial_{n+1} \maps C_{n+1}(X) \to C_n(X)$ is totally
unimodular (meaning every subdeterminant is in $\{ -1, 0, 1 \}$), the
OHCP reduces to an ordinary linear programming (LP) problem, and those
\emph{can} be solved in polynomial time. This LP problem is the
integer program with the integrality constraints dropped, i.e., it is
the LP relaxation of the integer linear program. Total unimodularity
implies that the constraint polyhedron is integral~\cite{VeDa1968},
and thus so is the solution to the linear program.

There is a simple criterion for when $\partial_{n+1}$ is totally
unimodular.  Recall that a pure subcomplex of $X$ of dimension $k$ is
a union of $k$-simplices of $X$ including all their subsimplices.  We
say that $X$ is \emph{relatively torsion-free in dimension $n$} if
$H_n(L, L_0)$ is torsion-free for all pure subcomplexes $L_0 \subset
L$ of dimensions $n$ and $n+1$ respectively.  Examples include any
orientable manifold of dimension $n+1$, or any simplicial complex that
embeds in $\R^{n+1}$.  It turns out that $\partial_{n+1}$ is totally
unimodular if and only if $X$ is relatively torsion-free in dimension
$n$ and so:
\begin{theorem}[\cite{DeHiKr2010}]\label{thm-DeHiKr-main}
  If $X$ is relatively torsion-free in dimension $n$ then the OHCP for
  $a \in C_n(X)$ can be solved in polynomial time.
\end{theorem}

\begin{figure}[t]
  \begin{center}
  \includegraphics[scale=0.22]
  {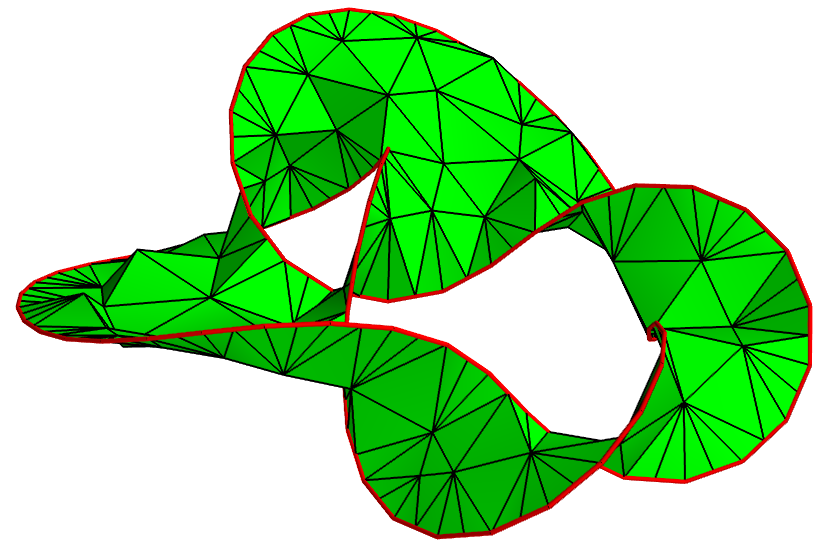}

  \vspace{2ex}

  \includegraphics[scale=0.22]
  {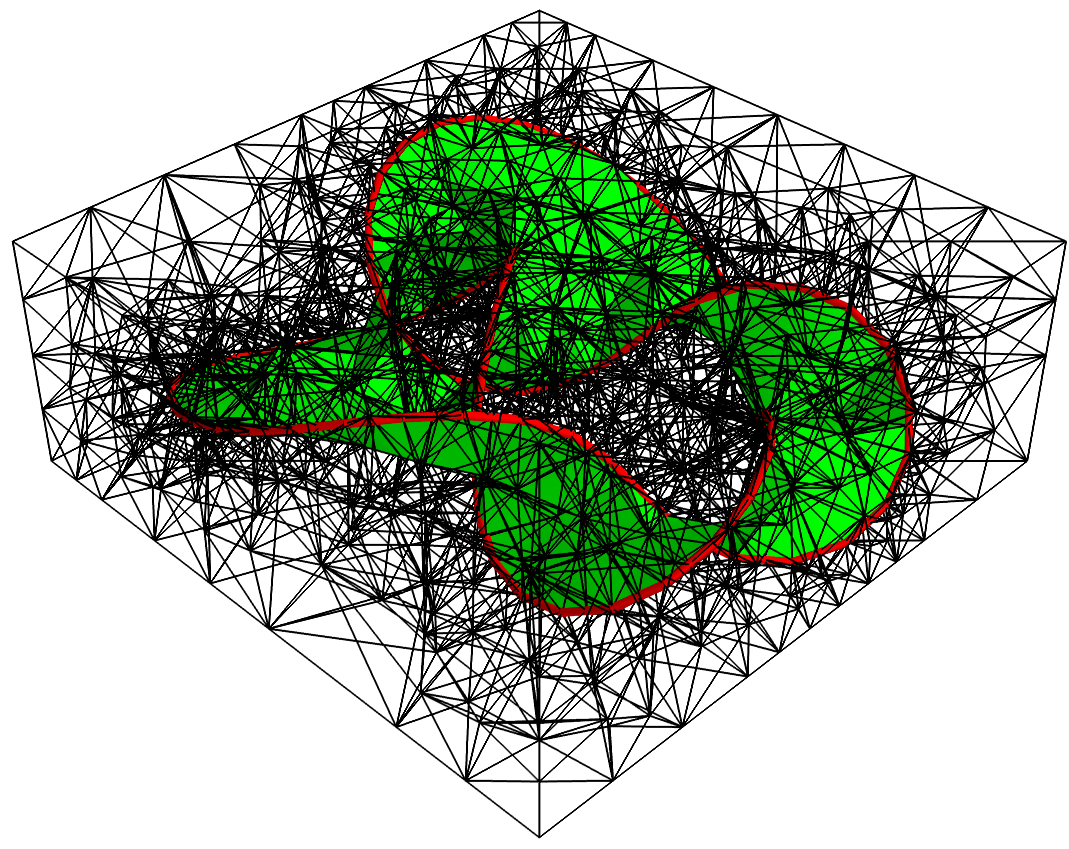}
  \end{center}
  \caption{At top is the solution (green) to the OBCP for the
    1-cycle that is given by the knot $5_2$ (red). The knot
    is a subcomplex of the 1-skeleton of a tetrahedral mesh of a
    cuboid shown in the lower picture.}
  \label{fig-5_2}
\end{figure}

Turning now to the Optimal Bounding Chain Problem (OBCP), assume that
instead we are given a lower dimensional chain $b \in C_{n-1}(X)$ and
we seek the minimum norm $c \in C_n(X)$ whose boundary is this $b$.
(Of course, if $[b] \in H_{n-1}(X)$ is nonzero this question is moot.)
Some examples are shown in Figures~\ref{fig-scherk} and \ref{fig-5_2}.
For certain $X$ we can relate these two problems:
\begin{theorem}\label{thm-OBCP-OHCP-equiv}
  Suppose $a \in C_n(X)$ is such that $b = \boundary_n a$. If $H_n(X)
  = 0$ then the OHCP (for that $a$) is identical to the OBCP (for
  that $b$).  Hence they have the same
  optimal solutions.
\end{theorem}
\begin{proof}
  In both problems, we seek a $c \in C_n(X)$ of minimal $\ell^1$-norm,
  so the claim is that the constraints on $c$ are actually the same in
  either case.  Since $H_n(X) = 0$, having $b = \partial_n c$ is
  equivalent to $c = a + \partial_{n+1} x$ as if we have the former
  then $\partial_n (c-a) = b - b = 0$ and thus there is an $x$ with
  $\partial_{n+1} x = c - a$.
\end{proof}

\begin{corollary} \label{thm-OBCP-P} Suppose $H_n(X) = 0$ and $X$ is
  relatively torsion-free in dimension $n$.  Then the OBCP for $b \in
  C_{n-1}(X)$ can be solved in polynomial time.  
\end{corollary}

\begin{proof}
  Set up the OBCP for $b$ as an integer LP, and then quickly solve its
  LP relaxation. If $b$ is not a boundary, then the feasible set is
  empty and we're done. Otherwise, we claim the solution is actually
  integral.  By Theorem \ref{thm-OBCP-OHCP-equiv} this LP is identical
  to one for the OHCP for some (uncomputed) $a \in C_n(X)$.  From our
  discussion of the proof of Theorem~\ref{thm-DeHiKr-main}, we know
  the latter LP has an integral solution as needed.
\end{proof}

\begin{remark}
  Consider a M\"obius strip embedded in $\R^3$. Enclose it in a cube
  and triangulate the cube with tetrahedra such that the strip is part
  of the 2-skeleton of the complex. Let $X$ be the cube triangulation
  and $b$ the 1-cycle carried by the topological boundary of the
  M\"obius strip. Here $\boundary_3$ is totally unimodular, which
  guarantees that $X$ is relatively torsion-free in dimension 2, as
  required by Cor.~\ref{thm-OBCP-P}. When we solve the OBCP problem as
  a relaxed LP, the constraints are $\partial_2 c = b$. The constraint
  matrix $\boundary_2$ is not totally unimodular
  \cite{DeHiKr2010}. Nevertheless, Cor.~\ref{thm-OBCP-P} guarantees
  that the minimizer $c$ will be integral. This does not contradict
  the result about equivalence of total unimodularity and
  integrality. That result says that total unimodularity is equivalent
  to the integrality of the constraint polyhedron for \emph{all}
  right-hand sides in the polyhedral constraint equations. Whereas in
  Cor.~\ref{thm-OBCP-P} we are saying that the solution is integral if
  the right-hand side $b$ is in the image of $\partial_2$ which
  certainly doesn't include all 1-chains in $X$.
\end{remark}

\subsection{Previous work on OBCP}
The OBCP in the trivial homology case as above has appeared in
Sullivan's thesis~\cite{SullivanThesis}, and in the work of
Grady \cite{Grady2010alt} and that of Gortler and his
coworkers~\cite{KiGo2004}.  When in addition $X$ is an $(n +
1)$-manifold, Sullivan gave a polynomial time algorithm for the OBCP
based on network flow. This idea also appears in~\cite{KiGo2004} and
related work. The basic idea in~\cite{KiGo2004} is that the cycle $b$
of dimension $n-1$ is on the boundary of the domain with trivial
$H_n(X)$. One introduces a source and a sink and connects them to the
centers of the top dimensional simplices on the boundary. These edges
are given infinite capacity. The dual graph of the codimension-1
skeleton then forms the rest of the edges in the network and these
edges have capacities equal to the volumes of the primal codimension-1
faces. Then by the maxflow-mincut theorem one obtains a maxflow and
hence an optimal chain.  It is an interesting question whether such
network flow methods can also be used to prove the more general
Cor.~\ref{thm-OBCP-P} which only requires that $X$ be relatively
torsion-free in dimension $n$.

\section{A relative version of the OBCP and a toy problem}
\label{sec-toy}

In geometric applications, one often cares not about the specifics of
the cycle $b \in C_{n-1}(X)$ but only its homology class in some
subcomplex $A \subset X$.  Before stating the problem in this context,
we recall the basics of relative homology.  The relative chain groups
are $C_n(X, A) = C_n(X)/C_n(A)$ which we also identify with the
submodule $C_n(X \setminus A)$ of $C_n(X)$.  The boundary maps for
$C_*(X, A)$ are induced from those of $C_*(X)$.  When $C_n(X, A)$ is
viewed as $C_n(X \setminus A)$, a relative cycle $c$ is simply one
where the support of $\partial_n c$ is contained in $A$.  Thus a
relative cycle gives rise to an element $[\partial_n c]$ in
$H_{n-1}(A)$ since $\partial_{n-1} \circ \partial_{n} = 0$.  (This map
from relative cycles to $H_{n-1}(A)$) is just the connecting
homomorphism in the long exact sequence of the
pair~\cite{Munkres1984}.)  We can now pose:
\begin{ROBCP}
  Let $A$ be a subcomplex of $X$, and $\beta \in H_{n-1}(A)$.  Find a
  relative cycle $c \in C_n(X, A)$ so that $[\partial_n c] = \beta$
  and $\norm{c}_1$ is minimal.
\end{ROBCP}

\begin{theorem}\label{thm-rel-OBCP}
  Suppose $H_n(X) = 0$ and $X$ is relatively torsion-free in dimension
  $n$.  Then the relative OBCP for $\beta \in H_{n-1}(A)$ can be
  solved in polynomial time.
\end{theorem}
\begin{proof}
  Suppose $\beta \in H_{n-1}(A)$ is specified by a cycle $b \in
  C_{n-1}(A)$.  First, we can quickly find a chain $c \in C_n(X)$ with
  $\partial_n c = b$ (or determine that none exists) either by solving
  a linear system over $\Z$~\cite{Dixon1982} or by solving this
  instance of the ordinary OBCP via Corollary~\ref{thm-OBCP-P}.

  Suppose $c' \in C_n(X)$ is any other chain with $\partial_n c' = b'$
  where $b' \in C_{n-1}(A)$ also represents $\beta$.  Since $H_n(X) =
  0$, it follows that $[c] = [c']$ in the relative homology group
  $H_n(X, A)$.  Thus, solving the relative OBCP for $\beta$ is the
  same as solving the OHCP for $C_n(X, A)$.  By the
  discussion in Section~\ref{sec:OBCP}, it is enough to show the
  \emph{relative} boundary map $\partial_{n+1} \maps C_{n+1}(X,A) \to
  C_{n}(X, A)$ is totally unimodular.  This is the case since its
  matrix is obtained by deleting certain rows and columns from that of
  the original $\partial_{n+1} \maps C_{n+1}(X) \to C_{n}(X)$.
\end{proof}

\subsection{Toy problem}\label{sec-toy-problem}

\begin{figure*}[tb]
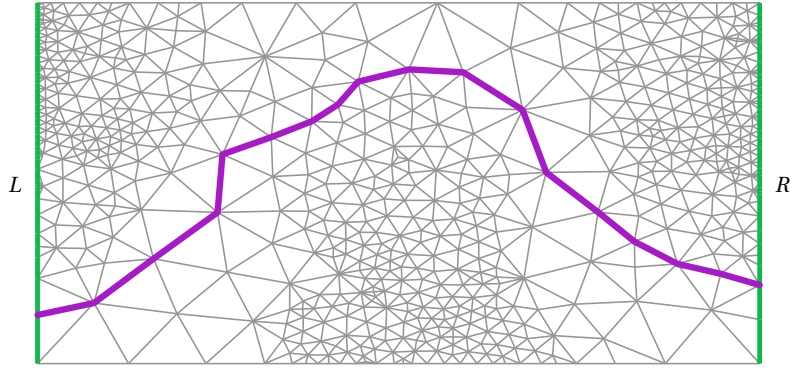

\begin{cxyoverpic}{(459,232)}{scale=0.6}{figs/knot/large-toy}
    ,(3,115)*++!R{L}
    ,(456,115)*++!L{R}
\end{cxyoverpic}
  \caption{A path which uses the fewest possible number of edges
  to join the two vertical sides.}
  \label{fig-toy}
\end{figure*}

We next give a quick application to a simple problem that nonetheless
has all the features of our proof of Theorem~\ref{thm-min-area} in
Section~\ref{sec-least-area}.  Let $X$ be a simplicial complex
homeomorphic to a square, and let $L$ and $R$ be subcomplexes
corresponding to a pair of opposite sides. We will show that
Theorem~\ref{thm-rel-OBCP} allows us to solve the following problem in
polynomial time:
\begin{goal}\label{prob-shortest}
  Find the shortest embedded simplicial path in the 1-skeleton of
  $X$ joining $L$ to $R$.
\end{goal} 
\noindent
See Figure~\ref{fig-toy} for an example.  
Of course, this problem can
be solved very efficiently by a variety of algorithms. We give it
primarily to introduce the idea of desingularization which is needed
for Theorem~\ref{thm-min-area}.  Let $A = L \cup R$.  Then $H_0(A) =
\Z \oplus \Z$ is generated by $[v_L]$ and $[v_R]$ for any vertices
$v_L$ and $v_R$ in $L$ and $R$ respectively.  Now $\beta = [v_R] -
[v_L]$ is $0$ in $H_0(X)$ and so consider relative cycles $c \in
C_1(X, A)$ with $[\partial_1 c] = \beta$.  Any embedded simplicial
path from $L$ to $R$ gives such a $c$, but of course not every such
$c$ comes from a path (e.g.~the coefficient on some edge could be
greater than $1$).  However, we will show that if $\norm{c}_1$ is
minimal then it does come from an embedded path.  This is necessarily
the path of minimal length, and hence we will have reduced
Problem~\ref{prob-shortest} to the relative OBCP for $\beta$.  Since
$H_1(X) = 0$, Theorem~\ref{thm-rel-OBCP} applies to let us quickly
find $c$, and hence answer Problem~\ref{prob-shortest}.
\begin{figure*}
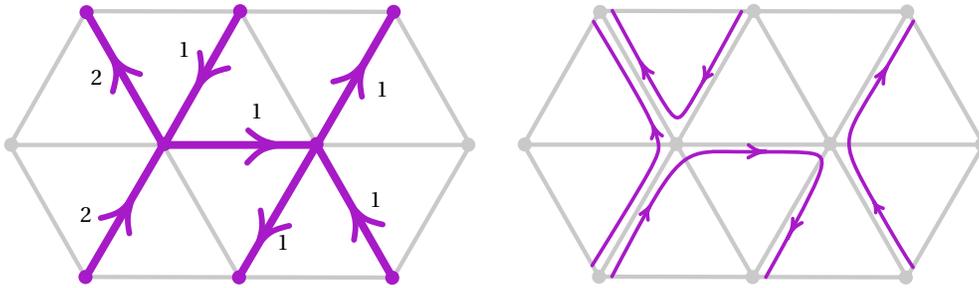

\begin{cxyoverpic}{(464,133)}{scale=0.8}{knot/desingularization}
    ,(45,33)*+!R{2}
    ,(50,98)*+!R{2}
    ,(95,111)*++!R{1}
    ,(119,71)*++!D{1}
    ,(125,27)*+!LU{1}
    ,(175,33)*+!D{1}
    ,(168,92)*++!L{1}
\end{cxyoverpic}
\caption{At left is part of a 1-cycle $c$ on a portion of a
  triangulated surface. At right is a desingularization of it. The
  resulting path need not be unique, but such a path will always
  exist. See Section~\ref{subsec-desingularization} for the higher
  dimensional desingularization.}
  \label{fig-desing}
\end{figure*}

Suppose $c$ is \emph{any} relative cycle in $C_1(X, A)$.  The
following desingularization procedure gives a collection of oriented
embedded loops and arcs which give the same class in $H_1(X,A)$.
Along an edge $e$ of $X$ where $c$ has coefficient $w$, we put
$\abs{w}$ strands parallel to $e$ oriented appropriately.  Then near
each vertex $v$ of $X \setminus A$ we connect up the strands without
introducing any crossings as shown in Figure~\ref{fig-desing}.
Because $\partial_1 c = 0$ at $v$, we can do this respecting the
orientations of the strands. That is, at the vertices, there is an
outgoing strand for every incoming one. Thus we can build a set of
loops and arcs homologous to $c$.

Now suppose $c$ minimizes $\norm{c}_1$ among the relative cycles with
$[\partial_1 c] = \beta$.  In the desingularization of $c$ there must
be at least one arc from $L$ to $R$, as for instance arcs joining $L$
to itself give $0$ in $H_0(A)$.  Slightly moving this arc, we push it
back onto the 1-skeleton $X^1$ of $X$ to give another relative cycle
$c' \in C_1(X,A)$ with $[\partial_1 c'] = \beta$.  The coefficient of
$c'$ on any edge is at most that of $c$, and so by minimality of
$\norm{c}_1$ we must have $c = c'$.  Thus $c$ corresponds to an
oriented path in $X^1$.  Moreover, this path must visit any given
vertex at most once, since otherwise a segment of the path forms a
closed loop which could be eliminated to reduce $\norm{c}_1$.  Thus
$c$ gives an embedded path from $L$ to $R$, as claimed.

\section{NP-completeness of the OBCP and the OHCP}
\label{sec-np-cmp}

In this section, we explain how the work of Agol, Hass, and Thurston
\cite{AgolHassThurston} shows that the decision problem variant of
OBCP is $\NP$-complete, and then use this to prove that the OHCP-D is
also $\NP$-complete.  Precisely, consider the following decision
problem:
\begin{obcpd}\label{obcpd}
  Given a simplicial complex $X$, a chain $b \in C_{n-1}(X )$, and an
  $L \in \N$, is there a chain $c \in C_n(X )$ with $\partial_n c = b$
  and $\norm{c}_1 \leq L$?
\end{obcpd}
\noindent
Here the complexity is in terms of the number of simplices in $X$ plus
the logs of $\norm{b}_1$ and $L$. We will show:
\begin{theorem}\label{thm-obcp-np}
  The OBCP-D is \textbf{NP}-complete.
\end{theorem}
\noindent
The proof of this is essentially contained in \cite{AgolHassThurston},
and indeed they use some clever tricks to reduce more geometric
problems like Knot Genus to the more combinatorial OBCP-D, though they
do not use the latter language explicitly.  Despite this, we include a
complete proof of Theorem~\ref{thm-obcp-np} as we need to modify the
construction to prove Theorem~\ref{thm-OHCP-NP}.  As a bonus, the
simpler context of the OBCP-D makes the idea of \cite{AgolHassThurston}
easier to digest for those not familiar with 3-manifold theory.

\begin{figure}[htb]
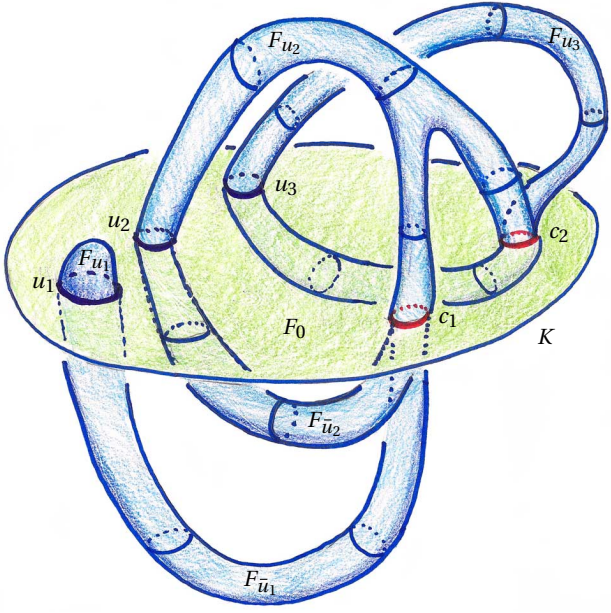

\begin{cxyoverpic}{(649,643)}{scale=0.24}{knot/AHT-complex-low}
   ,(275,30)*{F_{\ubar_1}}
    ,(342,197)*{F_{\ubar_2}}
    ,(101,368)*{F_{u_1}}
    ,(310,297)*{F_0}
    ,(300,596)*{F_{u_2}}
    ,(593,600)*{F_{u_3}}
    ,(47,343)*{u_1}
    ,(126,407)*{u_2}
    ,(301,444)*{u_3}
    ,(471,306)*{c_1}
    ,(587,395)*{c_2}
    ,(564,296)*!LU{K}
\end{cxyoverpic}
  \caption{The complex $X$ associated to $(u_2\vee \ubar_2 \vee
    \ubar_1) \wedge (u_2 \vee u_3 \vee \ubar_3)$. }
  \label{fig-AHT}
\end{figure}

One of two key ideas in \cite{AgolHassThurston} is the following
construction, which relates the OBCP-D to 1-in-3 SAT, which is
$\NP$-complete \cite{GareyJohnson1979, Schaefer1978}.  Recall that in
1-in-3 SAT, we are given boolean variables $U = \{u_1,\ldots, u_n\}$
and clauses $C = \{c_1, \ldots, c_m\}$, where each clause contains
three literals ($u_i$ or its negation $\ubar_i$) joined by $\vee$.
The question is whether there is a truth assignment for $U$ so that
each clause has \emph{exactly} one true literal.  We now build a
2-complex $X$ associated to an instance of 1-in-3 SAT by gluing
together several planar surfaces, that is, 2-spheres with (open) discs
removed.  Throughout this discussion, consult Figure~\ref{fig-AHT}
for an example.

The base surface $F_0$ has $n + m + 1$ boundary components, one
labeled by the symbol $K$ and the others by the elements of $U \cup
C$.  There is also a surface $F_{u}$ for each variable $u \in U$,
which has one boundary component labeled by $u$, and the others
labeled by $c \in C$ for each time $u$ (but not $\ubar$) occurs in
$c$. For the negation of each variable $u$ there is a surface
$F_{\ubar}$ with one boundary component labeled $u$ and the others
labeled by the appearances of $\ubar$ in the clauses. When only $u$ or
$\ubar$ appears in the clauses, then one of the surfaces is simply a
disc.

We triangulate each surface $F$ so that every boundary component
consists of three \one-simplices and the number of \two-simplices in
$F$ is $5 \abs{\partial F} - 4$, where $\abs{\partial F}$ denotes the
number of connected components of $\partial F$.  For each $F$, we fix
consistent orientations of its 2-simplices to create a relative cycle
which generates $H_2(F, \partial F) \cong \Z$; we denote the
corresponding chain in $C_2(F)$ by $[F]$.  Now we build $X$ by gluing
together all boundary components with the same labels, in such a way
that the gluings between $F_0$ (where every label appears) and any
$F_u$ or $F_\ubar$ is orientation reversing; in particular,
$\partial_2( [F_0] + [F_u])$ is $0$ along the circle labeled $u$.  We
let $b$ be the 1-cycle that corresponds to the boundary component of
$F_0$ that is labeled $K$, oriented so that it appears in
$\partial_2[F_0]$.  The key lemma is:
\begin{lemma}\label{lem-np}
  The 1-in-3 SAT instance $(U, C)$ has a solution if and only if there
  exists $d \in C_2(X)$ with $b = \partial_2 d$ and $\norm{d}_1 \leq
  1+ 6 n + 10 m$.
\end{lemma}

\begin{proof}
  To start, observe that if we denote the set of literals by $V = U
  \cup \setdefm{\big}{\ubar}{u \in U}$ then any solution $d$ to $b
  = \partial_2 d$ necessarily has the form
  \[
  d = [F_0] + \sum_{v \in V} k_v [F_v]  \quad \mbox{for some $k_v$ in $\Z$} 
  \]
  and the boundary of any such chain is supported on the labeled
  circles.  Since each 2-simplex lies in exactly one surface, the
  chains in the sum above have disjoint supports, and an easy
  calculation gives
  \[
  \norm{d}_1 = 1 + 5 n + 5 m + \norm{k_V}_1 + 5 \sum_{v
    \in V} \abs{k_v} m_v
  \]
  where $k_V$ is the vector $(k_v)$ and $m_v$ is the number of times
  $v$ appears in the clauses $C$.  Using $x_c, y_c, z_c$ to denote the
  literals that appear in a clause $c$, we can rewrite this as:
  \begin{equation}\label{eq-l1-norm-form}
  \norm{d}_1 = 1 + 5 n + 5 m + \norm{k_V}_1 + 5 \sum_{c
    \in C} \left( \abs{k_{x_c}} + \abs{k_{y_c}} + \abs{k_{z_c}} \right)
  \end{equation}

  To start the proof proper, first suppose we have a solution to the
  1-in-3 SAT instance, and let $d$ be the chain where $k_v$ is $1$ or
  $0$ depending on whether $v$ is true or false.    For each
  variable $u$, one of $k_u$ and $k_\ubar$ is $1$ and the
  other $0$; hence taking into account the contribution from $F_0$, we see
  that $\partial d$ is $0$ along the circle labeled $u$.  Now for a
  circle labeled by a clause $c$, as exactly one literal in $c$ is
  true we again see that $\partial d$ is $0$ along this circle.  Hence
  $\partial d = b$, and (\ref{eq-l1-norm-form}) gives that $\norm{d}_1
  = 1 + 6 n + 10 m$.  

  Conversely, let $d$ be a chain with $\partial d = b$ and $\norm{d}_1
  \leq 1 + 6 n + 10 m$.  As $\partial d$ is 0 along the circle labeled
  by $u$, it follows that at least one of $k_u$ and $k_\ubar$ is
  nonzero, and hence $\norm{k_V} \geq n$.  Similarly, for each clause
  $c$ at least one of $k_{x_c}, k_{y_c}, k_{z_c}$ must be nonzero to
  ensure $d$ has no boundary along the circle labeled $c$.  Hence from
  (\ref{eq-l1-norm-form}) and the bound on $\norm{d}_1$ it follows
  that $\norm{k_V} = n$ and each summand in the right-hand sum is $1$.
  Thus exactly one of $k_u$ and $k_\ubar$ is 1 and the other 0, and
  each clause has exactly one $F_v$ surface coming into it having
  nonzero weight.  Therefore $d$ corresponds to the needed solution to the
  1-in-3 SAT instance.
\end{proof}
\noindent
It is now easy to prove Theorem~\ref{thm-obcp-np} and then adapt this
construction to show Theorem~\ref{thm-OHCP-NP}. 
\begin{proof}[Proof of Theorem~\ref{thm-obcp-np}]
  The OBCP-D is in $\NP$ as we can use the cycle $c$ itself as the
  certificate.  One just has to check that $\partial c = b$, and this
  matrix vector multiplication is polynomial in the number of
  simplices in $X$. Conversely, the OBCP-D is $\NP$-hard since given
  an instance of 1-in-3 SAT, by Lemma~\ref{lem-np} there is an
  associated 2-complex $X$ (made from $O(n + m)$ simplices) and a
  1-cycle $b$ so that a solution to the 1-in-3 SAT problem is
  equivalent to finding $c$ with $\partial c = b$ and $\norm{c}_1 \leq
  1 + 6 n + 10 m$.
\end{proof}
\begin{proof}[Proof of Theorem~\ref{thm-OHCP-NP}]
  Let $CX$ be the cone on $X$, which is a 3-dimensional simplicial
  complex.  Let $Y$ be the 3-complex obtained from $CX$ by attaching a
  2-simplex $\sigma$ to the boundary component of $F_0 \subset
  X \subset CX$ labeled $K$.  For convenience, we use an $\ell^1$-norm
  on $C_2(Y)$ so that each 2-simplex in $X' = X \cup \sigma$ has
  weight 1, but the rest each have weight $10 + 6 n + 10 m$.  As $CX$
  is contractible, the space $Y$ is homotopy equivalent to $S^2$, and
  hence $H_2(Y) = \Z$.  Let $a \in C_2(Y)$ generate $H_2(Y)$ and have
  weight $+1$ on $\sigma$, where $\sigma$ is oriented compatibly with
  $F_0$.  We claim that our instance of 1-in-3 SAT has a solution if
  and only if $a$ is homologous to a cycle of weight at most $2 + 6n +
  10m$.  Such a cycle would have to be confined to $X'$, and thus have
  the form $\sigma + d$ where $d$ is as in the proof of
  Lemma~\ref{lem-np}, proving our claim and hence the theorem.
\end{proof}

\section{Least area surfaces bounded by a knot}
\label{sec-least-area}

Recall from Section~\ref{sec-intro} that a basic question about a
smooth knot $K$ in a closed Riemannian \3-manifold $Y$ is the minimal
area of an embedded surface $S$ with boundary $K$.  Agol, Hass, and
Thurston \cite{AgolHassThurston} considered the following discrete
version. Take $K$ to be a subcomplex of the 1-skeleton of a
triangulation of $Y$, where each simplex has a fixed geometric shape
corresponding to a simplex in $\R^3$ with rational edge lengths.  They
showed that the question of whether $K$ bounds a surface of area $\leq
A_0$ is $\NP$-hard.  Because they put no restriction on the surface
involved, it is not clear whether this question is in $\NP$.

Here, we consider another discretization which is a little more
combinatorial and will thus turn out to be $\NP$-complete.  For ease
of exposition, let us fix that $K$ is null-homologous in $Y$ as
otherwise there are no such $S$.  We switch focus to the exterior of
$K$, that is, the complement in $Y$ of a small open tubular
neighborhood of $K$.  This exterior is a compact \3-manifold whose
boundary is a torus.  Let $M$ be a simplicial complex triangulating
the exterior, where each 2-simplex has an ``area'' that is an
arbitrary natural number.  An orientable surface $S'$ in $Y$ with
boundary $K$ gives a properly embedded surface $S = S' \cap M$ in $M$
whose boundary generates the kernel $\pair{\lambda}$ of $H_1(\partial
M) \to H_1(M)$. (A properly embedded surface $S$ in $M$ is one such
that $\partial S$ is in $\partial M$.  A generator $\lambda$ for the
kernel of $H_1(\partial M) \to H_1(M)$ is a longitude curve on
$\partial M$.) Conversely, any properly embedded orientable surface
$S$ in $M$ where $[\partial S] = \lambda$ (in $H_1(\partial M)$) gives
a surface bounding $K$, after possibly adding some annuli and discs to
boundary components of $S$ to reduce the number of boundary components
to one.

To keep things combinatorial, we consider surfaces which lie in, or at
least near, the 2-skeleton of $M$.  Initially, we drop the condition
that the surfaces be embedded and consider the set $\cF$ of simplicial
maps $f \maps (S, \partial S) \to (M, \partial M)$ where $S$ is an
orientable surface with boundary and $f_*\big([\partial S]\big)$
generates the kernel of $H_1(\partial M) \to H_1(M)$. (Here $f_*$ is
the map at the level of homology induced by $f$ and $f_\#$, which
we will use later, is the induced map at the level of chains.) The
areas of the 2-simplices of $M$ can now be used to define the area of
this surface, which we denote $\Area (f)$.  We now consider:
\begin{minarea}\label{prob-LSA}
   Given $A_0 \in \N$ and the exterior $M$ of a null-homologous knot
   $K \subset Y$, is there an $f \in \cF$ with $\Area(f) \leq A_0$?
\end{minarea}
\noindent
With respect to the complexity of the number of simplices in $M$ and
 $\log A_0$, we show
\begin{theorem}\label{thm-LSA-hard}
  The Least Spanning Area Problem is \textbf{NP}-complete. 
\end{theorem}
\noindent
The proof that Least Spanning Area is $\NP$-hard is essentially the
same as in \cite{AgolHassThurston}, and that it is in $\NP$ will
follow from the desingularization procedure discussed below.

\subsection{Desingularization}\label{subsec-desingularization}

As in Section~\ref{sec-toy-problem}, a key tool is the following
well-known procedure for turning a relative cycle $c \in
C_2(M, \partial M)$ into a properly embedded surface $S$ representing
the same class in $H_2(M, \partial M)$.  Let $B$ be the union of small
balls about each vertex of $M \setminus \partial M$, and $T$ be the
union of $B$ with even smaller tubes about each edge of $M
\setminus \partial M$.  For each 2-simplex $\sigma$ in $M$, we take
oriented parallel copies of the hexagon $\sigma \setminus T$ according
to the weight of $c$ on $\sigma$.  (If some of the edges of $\sigma$
lie in $\partial M$, then $\sigma \setminus T$ may have fewer than six
sides.)  Now in the tube of $T$ about an interior edge $e$ of $M$, we
join the adjacent hexagons to form a properly embedded oriented
surface $S$ in $M \setminus B$; the picture here is analogous to the
product of Figure~\ref{fig-desing} with the interval, and we can
always do this because $\partial c = 0$ along $e$.  The surface $S$
meets the boundary of each ball $B_0$ in $B$ in a collection of simple
closed curves in the sphere $\partial B_0$.  We can take a disjoint
collection of disks in $B_0$ with the same boundary as $S
\cap \partial B_0$ and add them to $S$.  The result is a properly
embedded surface $S$ that is homologous to $c$.  The way we built it,
the surface $S$ has the following natural decomposition as a
simplicial complex so that the map that pushes it back onto the
2-skeleton of $M$ is simplicial.  In particular, we give $S$ the
simplicial structure where there is one triangle for each hexagon, one
edge for each gluing of hexagons across the tubes of $T$, and one
vertex for each disk added inside $B$.  The desired simplicial map $S
\to M$ just maps things to the corresponding simplices of $M$, e.g. a
triangle $\tau$ coming from a hexagon $h$ goes to the $\sigma$ that $h$
was build from.  We summarize our discussion as:
\begin{lemma}\label{lem-desing}
  Let $c$ be a relative cycle in $C_2(M, \partial M)$.  Then there is
  a simplicial surface $S$ and a proper embedding $S \to M$ that is
  arbitrarily close to a simplicial map $f \maps S \to M$.  Moreover,
  $\norm{c}_1 = \Area(f)$.
\end{lemma}
\noindent
We now use this to connect the Least Spanning Area Problem to the
decision problem variant of the relative OBCP: given $\beta \in
H_{n-1}(A)$ and $A_0 \in \N$ is there a relative cycle $c \in
C_n(X,A)$ with $[\partial_n c] = \beta$ and $\norm{c}_1 \leq A_0$?
\begin{theorem}\label{thm-LSA-OBCP}
  The Least Spanning Area Problem is equivalent to the relative OBCP-D
  for $M$ and $\lambda \in H_1(\partial M)$.
\end{theorem}
\noindent
Combining Theorem~\ref{thm-LSA-OBCP} with
Theorem~\ref{thm-rel-OBCP} immediately proves
Theorem~\ref{thm-min-area}, since orientable \3-manifolds are
relatively torsion-free in dimension 2.

\begin{proof}[Proof of Theorem~\ref{thm-LSA-OBCP}]

  Suppose that $c \in C_2(M, \partial M)$ solves the relative OBCP-D
  problem, i.e.~$[\partial c] = \lambda$ and $\norm{c}_1 \leq A_0$.
  Then by Lemma~\ref{lem-desing} there is a corresponding surface $f
  \in \cF$ with $\Area(f) = \norm{c}_1$.  

  Conversely, suppose $f \in \cF$ with $\Area(f) \leq A_0$.
  Consider $c = f_{\#}([S])$, which is a relative cycle in
  $C_2(M, \partial M)$ and moreover $[\partial c] = f_*\big([\partial
  S]\big) = \lambda$. Moreover $\norm{c}_1 \leq \Area(f)$ and so $c$
  solves this relative OBCP-D instance.
\end{proof}

\begin{proof}[Proof of Theorem~\ref{thm-LSA-hard}]
  First, the relative OBCP-D is in $\NP$ as we can just use $c$ as the
  certificate; the sizes of the coefficients are uniformly bounded by
  $A_0$ so this is small and checking that $[\partial c] = \lambda$ is
  a polynomial time computation in linear algebra over $\Z$,
  polynomial in the size of $M$.  Theorem~\ref{thm-LSA-OBCP} now gives
  that Least Spanning Area is in $\NP$.

  The argument that Least Spanning Area is $\NP$-hard is essentially
  the same as for the original discretization of the Riemannian least
  area problem studied in \cite{AgolHassThurston}.  Their proof uses a
  suitable \3-manifold built from the 2-complex of
  Section~\ref{sec-np-cmp}.  The only modification needed here is that
  we've set things up to require that $K$ is null-homologous in $Y$.
  This can be arranged by adding a disk with weight $10 + 6n + 10m$ to
  the boundary component of $F_0$ labeled $K$, and also adding there a
  small annulus where the unglued boundary becomes the new $K$.
\end{proof}

\begin{remark}
  The work of Sullivan in \cite{SullivanThesis} is the closest
  antecedent to our Theorem~\ref{thm-min-area}.  There, his motivation
  is rigorously approximating the area of a smooth least area surface
  in $\R^3$ bounding $K$, and the bulk \cite{SullivanThesis} is
  showing that certain types of meshes necessarily contain a solution
  $c$ to the OBCP for $K$ whose area is within the given tolerance of
  the minimal area.  While he gives a fast algorithm to find $c$, he
  does not insist that $c$ gives an embedded surface.  After all,
  the regularity theorems for least area surfaces guarantee that there
  is always such a surface of smaller area than $c$.  However, once
  one completely discretizes the problem as we have done here, a
  priori there could be a difference between the OBCP and the more
  geometric question about embedded surfaces. From the point of view
  of our desired application to Knot Genus as discussed in
  Section~\ref{sec-future}, it is important to have a concrete least
  area surface rather than just a homology class.
\end{remark}

\begin{remark}
  When the ambient manifold $Y$ is $\R^3$, $S^3$, or $\mathbb{H}^3$,
  an important alternate approach to finding least area surfaces was
  introduced by Pinkall and Polthier in \cite{PinkallPolthier1993},
  see also \cite{Polthier2005, HiPoWa2006}. They consider simplicial
  surfaces where the vertices are allowed to have arbitrary positions
  in $Y$.  A discrete minimal surface is then one for which small
  perturbations of the vertices do not decrease total area.  The paper
  \cite{PinkallPolthier1993} gives an algorithm that takes an initial
  surface $S_0$ bounded by $K$ and flows it toward a discrete minimal
  surface $S$.

  There are two problems with using the approach of
  \cite{PinkallPolthier1993} to solve Least Spanning Area in the
  restricted case of $Y = \R^3$.  The first is that singularities can
  develop during this flow \cite{PinkallPolthier1993}, and it seems
  unknown whether one still always ends up with a discrete minimal
  surface \cite[\S 5.3]{Polthier2005}.  A more fundamental problem is
  that there can be many discrete minimal surfaces spanning $K$ which
  are not least area, and any flow method can get stuck on such a
  surface depending on the choice of $S_0$.  An extreme case is the
  knots of \cite{Lyon1971} which have infinitely many incompressible
  spanning surfaces where no essential simple closed curve on the
  surface bounds a disk in $\R^3 \setminus K$. Each of these surfaces
  should be isotopic to a discrete minimal surface, leading to
  infinitely many distinct minimal spanning surfaces.

  To quickly approximate the smooth least-area surface spanning $K$
  in $\R^3$, a promising strategy is to first use
  Theorem~\ref{thm-min-area} with respect to some mesh containing $K$
  to produce $S_0$ and then apply \cite{PinkallPolthier1993} which is
  not constrained by the initial choice of mesh.
\end{remark}

\section{Future work} \label{sec-future}

The Knot Genus and Least Spanning Area problems have a very similar
flavor, and hence Theorem~\ref{thm-min-area} is compelling evidence
for the tractability of Conjecture~\ref{conj-knot-P}.  However, these
two problems are not always solved by the same surface --- one can
always cook up triangulations so that the least area surface is not
the one of minimal genus (a very striking example of this is
\cite{HaSnTh2003}).  Still, for the triangulations that one encounters
in practice, they should frequently be the same. Thus a natural place
to look for the minimal genus surface is the one constructed in proving
Theorem~\ref{thm-min-area}.

Here, it important to emphasize how much faster the method of
\cite{DeHiKr2010} is in practice compared to traditional normal
surface algorithms.  Using normal surfaces, a triangulation with 30 or
35 tetrahedra is near the limit of feasible computation, whereas
\cite{DeHiKr2010} can handle examples with more than 20{,}000
tetrahedra (see Figure~\ref{fig-scherk} below).  Thus we should be
able to work with reasonably fine triangulations of $M$, which could
increase the chance that the least area surface is minimal genus.  For
instance, from the Thurston/Perelman point of view, one could take
some combinatorial approximation to a hyperbolic metric on $M$.
Unfortunately, doing so will not completely eliminate the issue of
least area surfaces not having minimal genus; in the smooth category,
a folk theorem gives a hyperbolic \3-manifold with a homology class
whose minimal area representative is compressible
(cf.~\cite{Hass1992}).

However, going to a large number of tetrahedra raises a different
issue: we need to know the genus of the surface $S$ constructed from
the $\ell^1$-minimal cycle $c$.  In fact, the surface $S$ is not
unique, as there can be many choices for how to pair up the triangles
near the edges of $M$ (cf.~Figure~\ref{fig-desing}), sometimes
resulting in exponentially many such surfaces in terms of $\| c \|_1$.
Moreover, these choices affect how many disks we add near the vertices
of $M$ to complete the surface, and thus affect the Euler
characteristic and hence the genus.  This leads to the natural
question: Can one quickly determine $\min\big( - \chi(S) \big)$ over
all surfaces $S$ resulting from $c$?  It might also be the case that
the classes $c$ that one finds in practice don't have many surfaces
associated to them, and we will run computer experiments on this.

\subsection*{Acknowledgements} The authors were partially supported by
the US National Science Foundation under grant DMS-0707136 and NSF
CAREER grant DMS-0645604.  We thank the SoCG referees and Joshua
Dunfield for their extensive and very helpful comments on the original
version of this paper.  We also thank Joel Hass for extremely useful
correspondence, and Steven Gortler for pointing out the result in
Sullivan's thesis and for explaining his work on OBCP.

\bibliographystyle{acmurldoi} 
\bibliography{hirani,knotgenus}

\end{document}